\documentclass[submission, copyright, creativecommons]{eptcs}

\usepackage{preamble}

\providecommand{\event}{Proceedings of the Workshop on Coalgebra,\\Horn Clause Logic Programming and Types (CoALP-Ty'16)\\
Edinburgh, Scotland, UK, November 28-29, 2016}

\title{Structural Resolution for Abstract Compilation of Object-Oriented Languages}
\def\titlerunning{Structural Resolution for Abstract Compilation of Object-Oriented Languages}

\author{
    \begin{authors}{2}
    Luca Franceschini & Davide Ancona
    \institute{University of Genoa\\
    Italy}
    \email{luca.franceschini@dibris.unige.it & davide.ancona@unige.it}
    \end{authors}
\and
	Ekaterina Komendantskaya
	\institute{Heriot-Watt University\\
	Edinburgh, Scotland, UK}
	\email{e.komendantskaya@hw.ac.uk}
}
\def\authorrunning{L. Franceschini, D. Ancona \& E. Komendantskaya}

\begin{document}
\maketitle
\begin{abstract}
We propose \emph{abstract compilation} for precise static type analysis of object-oriented languages based on \emph{coinductive logic programming}. Source code is translated to a logic program, then type-checking and inference problems amount to queries to be solved with respect to the resulting logic program. We exploit a coinductive semantics to deal with \emph{infinite terms and proofs} produced by recursive types and methods. Thanks to the recent notion of \emph{structural resolution} for coinductive logic programming, we are able to infer very precise type information, including a class of irrational recursive types causing non-termination for previously considered coinductive semantics. We also show how to transform logic programs to make them satisfy the preconditions for the operational semantics of structural resolution, and we prove this step does not affect the semantics of the logic program.
\end{abstract}

\vspace*{-0.1in}
\section{Introduction}

Object-oriented programming is the most widespread computational paradigm in programming languages\footnote{See for instance Tiobe index at \url{http://www.tiobe.com/tiobe-index/}.}. 
Statically typed languages like Java, {C\texttt{\#}}, and C++ are heavily employed
for large scale software development in several domains;
however, for implementing applications based on Web or scientific programming, where often less skilled  developers are involved, 
dynamically typed object-oriented languages like JavaScript and Python have  gained more popularity for their gentler learning
curve, flexibility and ease of use.

Although such languages favour rapid development, and dynamic software adaptation,
they also lack the benefits of a static type system: since all errors are detected only 
at runtime, programs are less reliable, and debugging and testing are more challenging;
furthermore, the absence of type information is a severe obstacle to more efficient implementations,
and more effective IDE tools.  

Static typing needs more effort and knowledge from the programmer, who is burdened by the task of annotating code with types;
this load is bearable for simple nominal type systems, but when more accurate type analysis is required, one has to resort to structural type systems, which are more verbose, complex, and, thus, less intuitive to grasp;
for instance, Java wildcards are a form of structural type, though integrated with the nominal type system, which cannot be tamed so simply by ordinary programmers.

From the considerations above we can draw the conclusion that there exists
a fundamental trade-off between the benefits of static typing, and those of dynamic typing.
In order to reduce this ``gap'', type inference, and, more in general, any static type analysis
which does not require type annotations, is a viable solution;
programmers are relieved from declaring types, and still have 
all benefits of a dynamic language, but also an effective analysis tool 
to develop more reliable, and maintainable software.
Unfortunately, depending on the language in use and on the expressive power of
the type system, type analysis of a dynamic language can become quite hard (or even impossible,
i.e., undecidable) to solve. 

Consider the following program implementing linked lists,
written in a hypothetical dynamic object-oriented language; for simplicity we 
adopt a Java-like syntax, but the program contains no type annotations.

\begin{listing}[H]
\begin{minipage}{.4\textwidth}
\begin{minted}[fontsize=\footnotesize]{Java}
class EList extends Object {
    EList() {
        super();
    }
    addLast(elem) {
        new NEList(elem, this)
    }
}
\end{minted}
\end{minipage}%
\begin{minipage}{.6\textwidth}
\begin{minted}[fontsize=\footnotesize]{Java}
class NEList extends Object {
    head;
    tail;
    NEList(head, tail) {
        super();
        this.head = head;
        this.tail = tail;
    }
    addLast(elem) {
        new NEList(this.head, this.tail.addLast(elem))
    }
}
\end{minted}
\end{minipage}
\caption{Untyped linked lists. \texttt{EList.addLast} simply creates a new list with (only) the given element, while \texttt{NEList.addLast} recursively reach the end of the list to add the element, and returns the new list.}
\label{lst:lists}
\end{listing}
Depending on the expressive power of the underlying type system, a static analysis tool could be able or not to
successfully analyze the expression \mint{Java}{new EList().addLast(42).addLast(false).head} and 
compute its expected type \mintinline{Java}{int}; however, in a dynamically typed language a tool
rejecting such an expression would not be considered very useful, since it should be quite
natural for a dynamic language to allow manipulation of lists of heterogeneous 
elements. Therefore, dynamic languages call for more precise type analysis able to support
both \emph{parametric} and \emph{data polymorphism}; the former is the ability to pass arguments of unrelated types 
to the same parameter, the latter allows assignment of values of unrelated types to the same class field. 
Correct type analysis of the expression above requires parametric polymorphism because the same
method \mintinline{Java}{addLast} of class \mintinline{Java}{NEList} is invoked twice with the first argument of type
\mintinline{Java}{int} and \mintinline{Java}{boolean}, respectively, but also data polymorphism is needed, because
the two invocations of \mintinline{Java}{addLast} assign to the field \mintinline{Java}{head} values of
type \mintinline{Java}{int} and \mintinline{Java}{boolean}, respectively.


Supporting parametric and data polymorphism requires the use
of advanced structural types, and ensuring termination of the analysis in presence of recursive types and methods can be challenging. In this paper we investigate an improvement of \emph{abstract compilation} \cite{AnconaL09} to get more
precise type analysis of object-oriented code involving recursive method invocations. 

Abstract compilation is a modular approach to type analysis that exploits logic programming;
programs under analysis are abstractly compiled to logic programs, and then analysis is performed
through goal resolution.

For instance, in order to infer the type of the last expression about list classes we considered, a goal clause similar to the following one could be used:
\[ \goal{\mli{new}(\mli{elist}, [], E), \mli{invoke}(E, \mli{addLast}, [\mli{int}], R), \mli{invoke}(R, \mli{addLast}, [\mli{int}], R')} \]
The three atoms encode the calls to the constructor and to the \mintinline{Java}{addLast} method (twice), respectively.
Variables \(E, R, R'\) are the types resulting from the operation.
After formulating the goal query, it has to be resolved with respect to the logic program generated by abstract compilation of the original one, as it will be shown in Section \ref{sec:ac}.
Finally, the computed substitutions will give terms encoding types, thus effectively solving the inference problem.

To support parametric and data polymorphism in the presence of recursion,
the resolution method has to support the coinductive interpretation of the generated
logic program \cite{SimonMBG06} based on the greatest complete Herbrand model. However, implementation
of the operational semantics of coinductive logic programming (Co-LP for short) fails
to successfully analyze some kinds of recursion. 

In this paper we show how this drawback can be overcome by adopting \emph{structural resolution} \cite{KomendantskayaJ15} for the inference engine used for abstract compilation; thanks to the notion of \emph{productivity},
structural resolution allows an operational semantics which is more expressive than Co-LP (under certain assumptions that will be discussed).


The rest of the paper is organized as follows.
Section~\ref{sec:lp} introduces the necessary background on coinductive logic programming and presents structural resolution,
while Section~\ref{sec:ac} is a detailed introduction to abstract compilation.
Section~\ref{sec:ac-c} motivates the usefulness of structural resolution for
abstract compilation and shows how it can improve previous results, which is the main contribution of this work.
Section~\ref{sec:prod} is devoted to the notion of productivity, and shows a transformation technique that guarantees productivity for abstract compilation.
Section~\ref{sec:conclu} contains some concluding remarks and future work directions.


\section{(Coinductive) Logic Programming}
\label{sec:lp}
\subsection{Logic Programming Preliminaries}
Given a first-order signature consisting of variables, function and predicate symbols, we define terms inductively as is standard:
 they can be either \emph{variables} or \emph{function symbols} of arity \(n\) applied to \(n\) terms (\(f(t_1, \dots, t_n)\)). \emph{Constants} are function symbols of arity \(0\).
 \emph{Atoms} or atomic formulas have the shape \(p(t_1, \dots, t_n)\), where \(p\) is a \emph{predicate symbol} of arity \(n\) and \(t_1, \dots, t_n\) are terms.  
\emph{Logic programs} are finite sets of \emph{definite Horn clauses} (clauses for short) \(A \lpimpliedby B_1 \lpand \dots \lpand B_n\), where $A, B_1, \ldots B_n$ are atoms. \(A\) is called the \emph{head} of the clause and \(B_1 \lpand \dots \lpand B_n\) is called the \emph{body} of the clause. When the body is empty the head  is considered to be \(\mli{true}\).  A \emph{goal clause} has the shape \(\goal{A_1 \lpand \dots \lpand A_n}\).

A \emph{substitution} is a finite (partial) mapping from variables to terms, where all variables are simultaneously substituted.
Two terms $t_1$ and $t_2$ are \emph{unifiable} by substitution $\sigma$ if $\sigma(t_1) = \sigma(t_2)$; \(t_1\) \emph{matches} \(t_2\) by $\sigma$ if $\sigma(t_1) = t_2$. If in these two cases,  $\sigma$ is additionally the most general substitution, we say it is the \emph{most general unifier (mgu)} or \emph{most general matcher (mgm)} respectively.
Terms are said to be \emph{ground} if they contain no variables. Given a term \(t\), a substitution \(\sigma\) is \emph{grounding for \(t\)} if \(\sigma(t)\) is ground. All these definitions can be extended to atoms and clauses in the standard way.

Traditionally, the inductive semantics of logic programs is given by Herbrand models. The \emph{Herbrand base} \(B_P\) of a logic program \(P\) is the set of all ground atoms built from function and predicate symbols in \(P\). The \emph{least Herbrand model} \(M_P\) is the smallest subset of \(B_P\) that is also a model for each clause in \(P\). An atom \(A\) is \emph{logically entailed} from \(P\) if \(\sigma(A) \in M_P\), for some grounding substitution \(\sigma\) for \(A\).

Given a logic program \(P\) and a goal clause \(\goal{A_1 \lpand \dots \lpand A_n}\), \emph{SLD resolution} is a semi-decision procedure to check whether \(A_1, \dots, A_n\) are logically entailed from \(P\) and, if so, to compute a substitution \(\sigma\) such that \(\sigma'(\sigma(A_i)) \in M_P\) for every \(A_i\) in the goal, and for all substitutions $\sigma'$ grounding for $\sigma(\goal{A_1 \lpand \dots \lpand A_n})$. Thus, goal clauses can be seen as queries to be solved with respect to a logic program, and the computed substitution encodes the answer (if any).


\begin{definition}[SLD-resolution reduction]
If \(P\) is a logic program and \(\goal{A_1\lpand\dots\lpand A_n}\) is a goal clause, the \emph{SLD-resolution reduction} (with respect to \(P\)) is given by \(\pargoal{A_1\lpand\dots\lpand A_n}
\sld_P
\pargoal{\sigma(A_1)\lpand \dots \lpand\sigma(A_{i-1})\lpand \sigma(B_0)\lpand
\dots \lpand \sigma(B_m)\lpand \sigma(A_{i+1})\lpand \dots\lpand \sigma(t_n)}\) if \(A \lpimpliedby B_0 \lpand \dots \lpand B_m \in P\) and \(\sigma\) is the mgu of \(A_i\) and \(A\).
\end{definition}

A more constructive definition of the least Herbrand models can be given in terms of \emph{fixed points} of a suitable function. Given a logic program \(P\), the \emph{immediate consequence operator} \(T_P : \powerset{B_P} \to \powerset{B_P}\) is a function defined on the powerset of the Herbrand base as follows:
\[ T_P(S) = \setbuild{A}{A \lpimpliedby B_1 \lpand \dots \lpand B_n \text{ ground instance of a clause in \(P\)}, \{ B_1, \dots, B_n \} \subseteq S} \]
Since \(T_P\) is monotonic for any logic program \(P\), by the Knaster-Tarski theorem it has a \emph{least fixed point}, which is precisely the least Herbrand model \(M_P\).

For the rest of the paper, we use the following syntactic conventions: function and predicate symbols start with a lowercase letter, and constants are sometimes numbers; variables start with an uppercase letter; atoms, clauses and logic programs are written as single uppercase letters.

\subsection{Coinduction in Logic Programming}
In inductive logic programming, only terminating SLD derivations are meaningful. However, there are logic programs for which there are no terminating derivations and there is no natural inductive semantics, yet they can still be understood \emph{coinductively}.

\begin{example}
\label{ex:zeros}
The following logic program \(P_{zeros}\) defines the infinite list of zeros:
\[ \mli{zeros}(\mli{cons}(0, X)) \lpimpliedby \mli{zeros}(X) \]
\end{example}

Starting from the goal \(\goal{\mli{zeros}(X)}\), SLD resolution does not terminate. Indeed, the program above has no inductive meaning, and its least Herbrand model is empty. Still, its clause has a clear meaning.

Recall that in the inductive interpretation, models contain only finite terms. Coinductive interpretation admits both finite \emph{and infinite} terms. Given a logic program \(P\), \(B_P^{co}\) is the \emph{complete Herbrand base} containing all finite and infinite atoms built on the top of function and predicate symbols in \(P\). For the coinductive interpretation, the \emph{greatest complete Herbrand model} \(M_P^{co}\) is considered, that is, the greatest subset of \(B_P^{co}\) that is also a model for \(P\). In example \ref{ex:zeros}, \(M_P = \emptyset\) but \(M_P^{co} = \{\mli{zeros}(\mli{cons}(0, \mli{cons}(0, \dots))\}\). 

The duality between the inductive and the coinductive interpretation extends to the fixed point semantics: inductive models are the least fixed point of the immediate consequence operator while coinductive models are the \emph{greatest} fixed point of the operator (extended to possibly infinite terms). The existence of the greatest fixed point is again ensured by the Knaster-Tarski theorem.

In the 80s, the notion of formulas \emph{computable at infinity} was introduced~\cite{Lloyd87}. An infinite formula  $A$ is computable at infinity, if there exists a finite formula $A'$ such that $A'$ has an infinite (and fair) SLD-derivation, and substitutions $\sigma_0,\sigma_1, \dotsc$ computed in the course of this derivation yield $\sigma_0(\sigma_1(\dots(A')\dots))$ $=$ $A$. For example, $\mli{zeros}(\mli{cons}(0, \mli{cons}(0, \dots))$ is computable at infinity for the program $P_{zeros}$ and the query $\mli{zeros}(X)$. 
In such cases we also say that the infinite SLD-derivation for $\mli{zeros}(X)$ is \emph{globally productive}, in a sense of producing an infinite term as a substitution.

Operationally, dealing with infinite terms and non-terminating derivations is  a challenge.
\emph{Co-LP} \cite{SimonMBG06} extends SLD resolution with a cycle detection mechanism that allows the derivation to be concluded when a goal unifies with a previously encountered one. Considering again Example \ref{ex:zeros}, the reduction \(\goal{\mli{zeros}(X)} \sld_{P_{zeros}} \goal{\mli{zeros}(X')}\) holds with the computed substitution \(\{X \mapsto \mli{cons}(0, X')\}\).
At this point, Co-LP checks if the two goals unify, and indeed they do:  the computed answer is \(\{X \mapsto \mli{cons}(0, X)\}\)\footnote{In the coinductive setting the occurs check needs to be removed.} which corresponds to the infinite term specified by the recursive equation \(X = \mli{cons}(0, X)\), that is \(\mli{cons}(0, \mli{cons}(0, \dots))\).

Note that the recursive term \(\mli{cons}(0, \mli{cons}(0, \dots))\) is \emph{regular}~\cite{Courcelle83} (a.k.a\ rational or cyclic) since it has a finite number of subterms, namely \(0\) and itself. Because  Co-LP's algorithm relies on unification of the looping coinductive goals, it can only handle regular terms and derivations. As a result, it does not terminate on irregular derivations, thus it is only sound but not complete w.r.t\ the greatest complete Herbrand model.

\subsection{Structural Resolution}
Structural resolution~\cite{KPS12-2,JKK15,KJS17} (or S-resolution for short) proposes a solution for cases when formulas computable at infinity are not regular. Consider the following example.
\begin{example}
\label{ex:from}
The coinductive program \(P_{\mli{from}}\) below has the following single clause:
\[\mli{from}(X, \mli{scons}(X,Y)) \lpimpliedby \mli{from}(s(X),Y) \]
Given the query \(\goal{\mli{from}(0, X)}\), and writing
$[\_|\_]$ as an abbreviation for the stream constructor
$\mli{scons}$, here we have that the infinite atom $a =
\mli{from}(0,[0|[s(0)|[s(s(0))|\dots]]])$ is computable at
infinity by $P_{\mli{from}}$ and it is also contained in the greatest complete Herbrand model of $P_{\mli{from}}$.  Coinductive reasoning on this query cannot be handled
by the loop detection mechanism of Co-LP because the atom $a$ is
irrational and the looping subgoals will fail to unify.
\end{example}

In such cases, it may still be possible to automatically prove that the SLD-derivation for the query $ \goal{ \mli{from}(0, X)}$
will be infinite, non-failing, and moreover will compute an infinite term at infinity, even if we cannot generate its closed form, as for $P_{zeros}$. 
In the core of this new argument is the detection of a regular pattern -- a constructor -- that works as a building block of the infinite term computed at infinity; in $P_{from}$ this constructor is $\mli{scons}$. We now explain the method that detects such patterns in S-resolution.

S-resolution~\cite{KPS12-2,JKK15,KJS17} stratifies the SLD-derivation steps into those done by term-matching and those requiring full unification. Term-matching in this case plays a role that pattern-matching on constructors of data structures plays in functional programming.


\begin{definition}\label{def:Sres}
\cite{FK16}\ If $P$ is a logic program and and $\goal{A_1 \lpand \dots \lpand A_n}$ is a goal clause, then:
\begin{itemize}


\item \emph{rewriting reduction}:
$\pargoal{A_1 \lpand \dots \lpand A_i \lpand \dots \lpand A_n}
\rewriting_P
\pargoal{A_1 \lpand \dots \lpand A_{i-1} \lpand \sigma(B_0) \lpand \dots \lpand \sigma(B_m) \lpand A_{i+1} \lpand \dots \lpand A_n}$ if $A \lpimpliedby B_0 \lpand \dots \lpand B_m \in P$ and \(\sigma\) is the mgm for \(A\) against \(A_i\) ($\sigma(A) = A_i$);
\item \emph{substitution reduction}:
$\pargoal{A_1 \lpand \dots \lpand A_i \lpand \dots \lpand A_n}
\substitution_P
\pargoal{\sigma(A_1) \lpand \dots \lpand \sigma(A_i) \lpand \dots  \lpand \sigma(A_n)}$ if $A \lpimpliedby B_0 \lpand \dots \lpand B_m \in P$ and $A_i$ and $A$ are unifiable via mgu $\sigma$.
\end{itemize}
The \emph{S-resolution reduction} with respect to $P$ is $\substitution_P \circ \rewriting^\mu_P$.
We write $\goal{A_1\lpand\dots\lpand A_n} \rewriting^\mu_P$ to
indicate the reduction of $\goal{A_1\lpand\dots\lpand A_n}$ to its $\rewriting$-normal
form with respect to $P$ if this normal form exists, and to indicate
an infinite reduction of $\goal{A_1\lpand\dots\lpand A_n}$ with respect to $P$ otherwise.
\end{definition}

One can show that under certain conditions, SLD-resolution reductions and structural resolution reductions are equivalent, see~\cite{FK16,KomendantskayaJ15}; but S-resolution has one advantage: it helps to detect the constructors from which the infinite data structure is built.
	


Firstly, we represent the structural resolution reductions as tree rewriting: 
the figure below shows how
 rewriting reduction steps can be represented as \emph{rewriting trees} and substitution reduction steps shown
horizontally as rewriting tree transitions. This separation makes it easy to see that in this derivation, the same pattern $[\_|\_]$ gets consumed by rewriting steps and 
gets added, or produced, in the substitution steps:
%
%
\begin{center}
\begin{tikzpicture}[scale=0.30,baseline=(current bounding box.north),grow=down,level distance=20mm,sibling distance=50mm,font=\footnotesize]
  \node { $\mli{from}(0, X)$};
  \end{tikzpicture}
$\stackrel{\{X \mapsto [0 \vert X']\}}{\substitution}$
\begin{tikzpicture}[scale=0.30,baseline=(current bounding box.north),grow=down,level distance=20mm,sibling distance=60mm,font=\footnotesize ]
  \node { $\mli{from}(0,[0 \vert X'])$}
          child { node {$\mli{from}(s(0),X')$}};
  \end{tikzpicture}
	$\stackrel{\{X'\mapsto [s(0) \vert X'']\}}{\substitution}$
	\begin{tikzpicture}[scale=0.30,baseline=(current bounding box.north),grow=down,level distance=20mm,sibling distance=60mm,font=\footnotesize ]
  \node { $\mli{from}(0,[0 \vert [s(0) \vert X'']])$}
	child { node{ $\mli{from}(s(0),[s(0) \vert X''])$ }
          child { node {$\mli{from}(s(s(0)),X'')$}}};
  \end{tikzpicture}~
$\stackrel{\{X'' \mapsto [s(s(0)) \vert X''']\}}{\substitution}$
\end{center}


We can only detect this pattern if rewriting trees are finite, i.e. all rewriting reductions are normalising.
By definition, \emph{observational productivity} of an S-resolution reduction for
a program $P$ and a query $A$ is in fact a conjunction of two properties~\cite{KJS17}:
\begin{itemize}
\item \emph{universal observability}: normalisation of all rewriting reductions in this S-resolution reduction, and
\item \emph{existential liveness}: non-termination of this S-resolution reduction.
\end{itemize}

S-resolution terminates when these two properties are satisfied.
Otherwise, S-resolution generates infinite derivations lazily, showing only partial answers.
For example, S-resolution will terminate lazily after the three substitution steps shown above, and will output the partial answer: $X = [0 \vert [s(0) \vert [s(s(0)) \vert X''']]]$.

One can show that observational productivity implies global productivity~\cite{KomendantskayaJ15,KJS17}.
Note that universal observability is thus a formal pre-condition for reasoning about observational productivity of S-resolution.
Not every non-terminating program is globally and observationally productive.


\begin{example}
\label{ex:infinitelp}
Consider the following logic program \(P\):
\begin{align*}
    p(f(X)) &\lpimpliedby p(X) \\
    q(X) &\lpimpliedby q(X)
\end{align*}
The two goal clauses \(p(X)\) and \(q(X)\) lead to the following non-terminating SLD derivations, respectively:
\begin{alignat*}{4}
    p(X) &\;\sld_P^{X=f(X')}\; &p(X') &\;\sld_P^{X' = f(X'')}\; &p(X'') &\;\sld_P^{X'' = f(X''')}\; &&\dotsb \\
    q(X) &\;\sld_P\; &q(X) &\;\sld_P\; &q(X) &\;\sld_P\; &&\dotsb
\end{alignat*}
\end{example}

In the first derivation, each derivation step gives a better approximation of the rational term \(f(f(\dots))\) by incrementally instantiating free variables 
\(X\), \(X'\), \(X''\), \dots. In the second one, the goal never changes, and there is no ``real'' progress. Nevertheless, both the rational atoms \(p(f(f(\dots)))\) and \(q(f(f(\dots)))\) belong to the greatest complete Herbrand model of the program above. 

Following the idea of \emph{computations at infinity} \cite{EmdenA85, Lloyd87}, only the first derivation actually computes an infinite term after an infinite number of steps and only the first derivation is globally productive. To see what happens with observational productivity, note that the clause \( q(X) \lpimpliedby q(X) \) makes the program break this requirement: \( q(X) \rewriting q(X) \rewriting \dotsb\,\). On the other hand, S-resolution reductions for \( p(X)  \) will be productive (again note that the constructor $f$ gets added in substitution reductions and consumed in rewriting reductions):
\[
    p(X) \; \substitution^{X=f(X')} \; p(f(X')) \; \rewriting \; p(X') \; \substitution^{X'=f(X'')}\; p(f(X'')) \; \rewriting \; p(X'') \dotsb 
 \]

Thus, S-resolution will work for queries on $p(X)$ but not $q(X)$. For the above derivation for $p(X)$, it will detect that $f$ plays a role of an infinite data structure constructor. 




\newcommand{\nullty}{\ensuremath{\mli{null}}}
\newcommand{\intty}{\ensuremath{\mli{int}}}
\newcommand{\boolty}{\ensuremath{\mli{bool}}}
\newcommand{\obj}[2]{\gobj{#1}{[#2]}}
\newcommand{\gobj}[2]{\ensuremath{\mli{obj}(\mli{#1},{#2})}}

\section{Abstract Compilation}
\label{sec:ac}

Abstract compilation \cite{AnconaL09,ACLD10-FoVeOOS10} is a technique developed in the context of 
object-oriented programming to exploit the potentialities of logic programming for supporting advanced static type analysis, as investigated also by other authors 
\cite{SulzmannStuckey08,HanVezzosi16}.

In a nutshell, the approach consists in translating the program under analysis into a logic program which abstracts
the semantics of the source program; then, performing static type analysis on the program amounts to
solving a goal w.r.t.\ the obtained logic program.

Abstraction is mainly obtained by structural types which represent set of values, following the semantic    
subtyping approach \cite{AnconaCorradi14,AnconaCorradi16}; boolean type constructors, as union types, and
record types allow quite precise analysis if employed in conjunction with abstract compilation; in particular, 
both parametric and data polymorphism are supported.
Solving a goal corresponds to symbolically executing the original source program
with types representing set of values.

Abstract compilation strives to reconcile compositional and global analysis, because
once the program under analysis has been abstractly compiled, its source code is no longer needed,
as long as it remains unmodified, and classes can be abstractly compiled separately. Since analysis corresponds to
goal solving, it can only be performed when the whole relevant program has been compiled; this limitation is also
a feature, because it promotes precise analysis through context sensitive data and parametric polymorphism.



Finally, abstract compilation offers interesting opportunities to fruitfully exploit compiler technology
\cite{AL-RAIRO11,AL-TCS12,AnconaCorradiFTfJP16} for more precise and efficient analysis.

\subsection{Abstract Compilation at Work} 
Let us consider the two classes implementing linked lists defined in listing~\ref{lst:lists} and show
how they could be translated into a logic program to perform type analysis on them.

The translation depends on the way values are abstracted, that is, the underlying type system.
In this particular case we may use the primitive types \nullty, \intty, and \boolty\
to represent the singleton value \mintinline{Java}{null}, and the sets of integer and boolean values, respectively;
then, \obj{c}{f_1{:}\tau_1,\ldots,f_n{:}\tau_n} represents the set of all instances created from class $c$ having
at least fields $f_1,\ldots,f_n$ associated with values of types $\tau_1,\ldots,\tau_n$, respectively. To make
the type system more expressive, we also introduce union types, corresponding to logical disjunction:
$\tau_1\lor\tau_2$ represents the set of all values which have type $\tau_1$ or $\tau_2$.

Types represent sets of values and are the terms manipulated by the generated logic programs; for instance, 
referring to the classes of listing~\ref{lst:lists}, the type $\obj{elist}{\ }\lor\obj{nelist}{head{:}\intty,tail{:}\obj{elist}{\ }}$
represents all objects implementing integer linked lists of length $\leq 1$.  

Predicates are introduced for representing the different kinds of declarations and constructs
of the source language. For instance, predicates $\mli{new}$ and $\mli{invoke}$ abstract object creation, and method
invocation, respectively, while predicate $\mli{hasmeth}$ represents method declarations. Consequently,
the atom $\mli{new(nelist,[\intty,\nullty],\obj{nelist}{head{:}\intty,tail{:}\nullty})}$ formally states that
the invocation of the constructor of class \mintinline{Java}{NEList} with arguments of type $\intty$ and $\nullty$, respectively,
returns a value of type $\obj{nelist}{head{:}\intty,tail{:}\nullty}$.
As another example, $$\mli{invoke(\obj{elist}{\ },addlast,[\intty],\obj{nelist}{head{:}\intty,tail{:}\obj{elist}{\ }})}$$
formally states that the invocation of method \mintinline{Java}{addLast} on an object of type $\obj{elist}{\ }$, and
argument of type $\intty$, returns an object of type  $\obj{nelist}{head{:}\intty,tail{:}\obj{elist}{\ }}$.

There exist two separate kinds of Horn clauses\footnote{We use Prolog syntactic conventions: variables start with an upper case letter, constants starts with a lower case letter and \([a, b, c, \dots]\) denotes a list. Moreover, we use the infix notation for the binary function symbol \(\lor\).} that are generated by the translation:
those encoding the abstract semantics of the source programming languages, which are independent of any analyzed program, and
those which are directly derived from the code under analysis. For instance, the clause
\begin{dmath*}
\mli{invoke}(\gobj{C}{F},M,A,R) \lpimpliedby \mli{hasmeth}(C,M, [\gobj{C}{F}|A],R)
\end{dmath*}
partly specifies\footnote{Two more clauses are needed to deal with union types, and with inherited methods.} the abstract semantics of method 
invocation; it states that the invocation of method $M$ on an object of type $\gobj{C}{F}$ with arguments of type $A$ returns a value of type
$R$, if the class $C$ of the receiver object has a method $M$ returning a value of type $R$ when invoked on
object \mintinline{Java}{this} of type $\gobj{C}{F}$ with arguments of type $A$. Thus, the semantics of $\mli{hasmeth}$ depends
on the code of the declared methods. Indeed, for each method declaration, a corresponding clause for predicate  $\mli{hasmeth}$ is generated;
for instance, the following clause is derived from the declaration of method \mintinline{Java}{addLast} in class \mintinline{Java}{EList}:
\begin{dmath*}
\mli{hasmeth}(\mli{elist},\mli{addlast}, [\mli{This},\mli{Elem}],R) \lpimpliedby \mli{new}(\mli{nelist},[\mli{Elem},\mli{This}],R)  
\end{dmath*}
It states that class \mintinline{Java}{EList} has method \mintinline{Java}{addLast} that, when invoked on the object  \mintinline{Java}{this} of type $\mli{This}$ with argument of type $\mli{Elem}$, returns a value of type $R$, providing that constructor of class
\mintinline{Java}{NEList} returns a value of type $R$ when invoked on arguments of type $\mli{Elem}$, and $\mli{This}$, respectively.

Analogously, the following clause is generated from the declaration of method \mintinline{Java}{addLast} in class \mintinline{Java}{NEList},
where predicate $\mli{fieldacc}$ abstracts the semantics of field access:
\begin{dmath*}
\mli{hasmeth}(\mli{nelist},\mli{addlast}, [\mli{This},\mli{Elem}],R) \lpimpliedby \mli{fieldacc}(\mli{This},\mli{head},H)
\lpand \mli{fieldacc}(\mli{This},\mli{tail},T)
\lpand \mli{invoke}(T,\mli{addlast},[\mli{Elem}],N)
\lpand \mli{new}(\mli{nelist},[H,N],R)  
\end{dmath*}

\subsection{Examples of Queries and Recursive Types}
We start by showing a simple goal to typecheck expression
\mintinline[escapeinside=||]{Java}{new EList().addLast(|$i$|)}, under the assumption that $i$ has type \mintinline{Java}{int}. This can be achieved by solving the goal 
\[
\goal{\mli{new}(\mli{elist},\lpemptylist,R) \lpand \mli{invoke}(R,\mli{addlast},[\intty],T)}
\]
which succeeds, as expected, with answer: 
\begin{eqnarray*}
R&=&\obj{\mli{elist}}{\ }\\
T&=&\obj{\mli{nelist}}{\mli{head}{:}\intty,\mli{tail}{:}\obj{\mli{elist}}{\ }}
\end{eqnarray*}
As a more elaborated example, let us consider the expression \mintinline[escapeinside=||]{Java}{new NEList(|$b$|,|$l$|).addLast(|$i$|)},
under the assumption that $b$, $l$, and $i$ have type $\boolty$, $\obj{\mli{elist}}{\ }$, and $\intty$, respectively; typechecking
this expression corresponds to solving the goal 
\[
\goal{\mli{new}(\mli{nelist},[\boolty,\obj{\mli{elist}}{\ }],R) \lpand \mli{invoke}(R,\mli{addlast},[\intty],T)}
\]
which succeeds for 
\begin{eqnarray*}
R&=&\obj{\mli{nelist}}{\mli{head}{:}\boolty,\mli{tail}{:}\obj{\mli{elist}}{\ }}\\
T&=&\obj{\mli{nelist}}{\mli{head}{:}\boolty,\mli{tail}{:}\obj{\mli{nelist}}{\mli{head}{:}\intty,\mli{tail}{:}\obj{\mli{elist}}{\ }}}.
\end{eqnarray*}
This example shows that typechecking can succeed also for expressions which build heterogeneous lists.

For a simple example of type inference, let us consider the problem of finding a valid type assignment for variables $x$ and $y$
to make the expression \mintinline[escapeinside=||]{Java}{|$x$|.addLast(|$y$|)} well-typed; this corresponds to the goal
\begin{dmath*}
\goal{\mli{invoke}(X,\mli{addlast},[Y],T)}
\end{dmath*}
which, for instance, succeeds for
\begin{eqnarray*}
X&=&\gobj{\mli{elist}}{F}\\
T&=&\obj{\mli{nelist}}{\mli{head}{:}Y,\mli{tail}{:}\obj{\mli{elist}}{F}}
\end{eqnarray*}
The fact that the logical variable $Y$ is not in the domain of the computed substitution means that any type can be safely assigned to 
$y$.
 
In the previous examples we have only considered types specifying linked lists of fixed length, but for
building more interesting types, recursion is needed; this is achieved by considering rational terms (a.k.a. regular or cyclic).
For instance, the unique term defined by the solution of the unification problem
\begin{dmath*}
T = \obj{\mli{elist}}{\ }\lor\obj{\mli{nelist}}{\mli{head}{:}\intty,\mli{tail}{:}T}
\end{dmath*}
corresponds to the recursive type specifying the set of all integer linked lists of arbitrary length. 
 
All example queries considered so far can be solved w.r.t. the standard inductive interpretation of Horn clauses, that is,
the least Herbrand model, even though method \mintinline{Java}{addLast} is recursive in class \mintinline{Java}{NEList}; however, if recursive types
are involved in queries, then the least Herbrand model is no longer sufficient to capture their intended meaning, as shown in
Section~\ref{sec:ac-c}.  



\section{Coinduction and Structural Resolution in Abstract Compilation}\label{sec:ac-c}

\subsection{The Need for Coinduction}

As already mentioned in the previous section, the intended meaning of goals and logic programs does not always coincide with the least Herbrand model. When recursive types and methods are involved, the coinductive interpretation is needed, i.e., the greatest complete Herbrand model has to be considered \cite{Lloyd87}.

Let us consider the recursive method \mintinline{Java}{replicate} from listing \ref{lst:replicate}: if
\mintinline{Java}{n} is not positive, then the method returns an empty list, otherwise it recursively builds a list of \mintinline{Java}{n-1} occurrences of \mintinline{Java}{x}, and then returns the (newly created) list where element \mintinline{Java}{x} has been added at the beginning. We refer to listing \ref{lst:lists} for the definitions of the classes \mintinline{Java}{EList} and \mintinline{Java}{NEList}.

\begin{listing}[ht]
\begin{minted}[escapeinside=||]{Java}
class ListFact extends Object {
    ListFact() { super(); }
    replicate(n, x) {
        if (n |\(\leq\)| 0) new EList()
        else new NEList(x, this.replicate(n-1, x))
    }
}
\end{minted}
\caption{Given an integer \mintinline{Java}{n} and an element \mintinline{Java}{x}, \mintinline{Java}{replicate} returns a list containing \mintinline{Java}{n} occurrences of \mintinline{Java}{x}.}
\label{lst:replicate}
\end{listing}

By means of abstract compilation, method \mintinline{Java}{replicate} would be translated\footnote{For the sake of readability, we have applied some simplifications to the resulting clause; however, such changes do not affect its semantics.} to the following clause:
\begin{dmath*}
\mli{hasmeth}(\mli{listfact}, \mli{replicate}, [\mli{This}, \mli{int}, X], E \lor \mli{NE})
\lpimpliedby \mli{new}(\mli{elist}, \lpemptylist, E)
\lpand \mli{invoke}(\mli{This}, \mli{replicate}, [\mli{int}, X], R)
\lpand \mli{new}(\mli{nelist}, [X, R], \mli{NE}) 
\end{dmath*}
The first atom of the body corresponds to the invocation of the constructor of  \mintinline{Java}{EList}, while the other two atoms are generated from the recursive invocation, and from the invocation of the constructor of  \mintinline{Java}{NEList}, respectively. Finally, the use of the conditional expression is reflected in the term \(E \lor \mli{NE}\). Note that the Horn clause above is the \emph{only} clause generated by abstract compilation for method \mintinline{Java}{replicate}. 

Let us now consider the expression \mintinline{Java}|new ListFact().replicate(10, 42)|.
To infer the type \(T\) of the expression above (w.r.t. the classes in listings \ref{lst:lists} and \ref{lst:replicate}), the following goal is generated: 
\[\goal{\mli{new}(\mli{listfact}, \lpemptylist, L) \lpand \mli{invoke}(L, \mli{replicate}, [\mli{int}, \mli{int}], T)}\]
However, such a goal fails to succeed if the inductive interpretation is considered; indeed,
the SLD derivation\footnote{Colours enlighten the substitution computed by unification along the way.} is \emph{non-terminating}:
\begin{gather*}
    \goal{\mli{new}(\mli{listfact}, \lpemptylist, \textcolor{red}{L}) \land \mli{invoke}(\textcolor{red}{L}, \mli{replicate}, [\mli{int}, \mli{int}], \textcolor{blue}{T})} \sld^* \\
    \goal{\mli{invoke}(\textcolor{red}{\mli{obj}(\mli{listfact}, \lpemptylist)}, \mli{replicate}, [\mli{int}, \mli{int}], \textcolor{blue}{T})} \sld^* \\
    \goal{\mli{hasmeth}(\mli{listfact}, \mli{replicate}, [\textcolor{red}{\mli{obj}(\mli{listfact}, \lpemptylist)}, \mli{int}, \mli{int}], \textcolor{blue}{T})} \sld^*\\
    \goal{\mli{hasmeth}(\mli{listfact}, \mli{replicate}, [\textcolor{red}{\mli{obj}(\mli{listfact}, \lpemptylist)}, \mli{int}, \mli{int}], \textcolor{blue}{\mli{obj}(\mli{elist}, \lpemptylist) \lor \mli{obj}(\mli{nelist}, [\mli{head} \colon \mli{int}, \mli{tail} \colon T'])})} \sld \dotsb
\end{gather*}
In the derivation above, every new atom for predicate $\mli{hasmeth}$ yields a better approximation for \(T\), but unfortunately the derivation never terminates. If we interpret \emph{coinductively} the logic program obtained by abstract compilation, thus considering its greatest complete Herbrand model rather than its least one, the goal above
is actually entailed by the program with a substitution instantiating \(T\) with the following type:
\[ \mli{obj}(\mli{elist}, \lpemptylist) \lor
\mli{obj}(\mli{nelist}, [\mli{head}\colon\mli{int}, 
    \mli{tail}\colon(\mli{obj}(\mli{elist}, \lpemptylist) \lor
    \mli{obj}(\mli{nelist}, [\mli{head}\colon\mli{int}, \dots ]
))]) \]
The type above represents all integer lists of arbitrary length; however, such a type corresponds to an infinite term.
Fortunately, there exists an equivalent type corresponding to the \emph{rational} term \cite{Courcelle83} specified by the following recursive equation:
\[ T = \mli{obj}(\mli{elist}, \lpemptylist) \lor \mli{obj}(\mli{nelist}, [\mli{head}\colon\mli{int}, \mli{tail}\colon T]) \]
This example shows that goals involving \emph{recursive} types and methods require
a coinductive interpretation of the logic program obtained by abstract compilation, in order to make static type analysis more precise.

As already illustrated in Section~\ref{sec:lp}, answer substitutions with rational terms can be computed by extending SLD resolution with cycle-detection techniques \cite{SimonMBG06}, as proposed with Co-LP; hence, an inference engine based
on Co-LP improves the result of static analysis performed with abstract compilation \cite{AnconaL09}. The Co-LP inference engine is however limited, since it succeeds with rational terms, and derivations, but it cannot handle more complex scenarios.

\subsection{Structural Resolution for Abstract Compilation}

Listing \ref{lst:buildlist} shows a slightly more involved example. Suppose we add to \mintinline{Java}{ListFact} a method \mintinline{Java}{buildList} that, given an integer \mintinline{Java}{n}, builds the list of integers \(1, 2, \dots, \mintinline{Java}{n}\) with an auxiliary method \mintinline{Java}{buildList} which exploits tail recursion with an accumulator parameter \mintinline{Java}{acc}; a more realistic Java implementation would of course avoid recursion, and use instead a simpler and more efficient loop, for which static type analysis for abstract compilation is less problematic if one exploits SSA intermediate form \cite{AL-RAIRO11,AL-TCS12,AnconaCorradiFTfJP16} during the compilation phase; however, in the past years object-oriented languages have begun to exploit
more and more patterns based on functional style programming, possibly with recursion and accumulators.
\begin{listing}[ht]
\begin{minted}[escapeinside=||]{Java}
buildList(n, acc) {
    if (n |\(\leq\)| 0) acc
    else this.buildList(n-1, new NEList(n, acc))
}
\end{minted}
\caption{Given an integer \mintinline{Java}{n}, \mintinline{Java}{buildList} returns the list of integers \(1, 2, \dots, \mintinline{Java}{n}\) followed by \mintinline{Java}{acc}, which is used as an accumulating parameter.}
\label{lst:buildlist}
\end{listing}

Abstract compilation of \mintinline{Java}{buildList} would yield\footnote{Again, for readability we are simplifying the clause that would be automatically generated.} the following Horn clause:
\[\resizebox{\linewidth}{!}{
\(\mli{hasmeth}(\mli{listfact}, \mli{buildlist}, [\mli{This}, \mli{int}, A], A \lor R) \lpimpliedby \mli{new}(\mli{nelist}, [\mli{int}, A], A') \lpand \mli{invoke}(\mli{This}, \mli{buildlist}, [\mli{int}, A'], R)\)
}\]
Suppose we want to infer the type $T_0$ of the expression
\mintinline{Java}{new List().buildList(42, new EList())}. Such an expression is abstractly compiled to the following goal:
\[\goal{ \mli{new}(\mli{listfact}, \lpemptylist, L) \lpand \mli{new}(\mli{elist}, \lpemptylist, A_0) \lpand \mli{invoke}(L, \mli{buildlist}, [\mli{int}, A_0], T_0)}
\]
The derivation for such a goal is again infinite, hence the coinductive interpretation is needed again:
\begin{gather*}
    \goal{ \mli{new}(\mli{listfact}, \lpemptylist, \textcolor{red}{L}) \lpand \mli{new}(\mli{elist}, \lpemptylist, \textcolor{blue}{A_0}) \lpand \mli{invoke}(\textcolor{red}{L}, \mli{buildlist}, [\mli{int}, \textcolor{blue}{A_0}], \textcolor{OliveGreen}{T_0}) } \sld^* \\
    \goal{ \mli{new}(\mli{elist}, \lpemptylist, \textcolor{blue}{A_0}) \lpand \mli{invoke}(\textcolor{red}{\mli{obj}(\mli{listfact}, \lpemptylist)}, \mli{buildlist}, [\mli{int}, \textcolor{blue}{A_0}], \textcolor{OliveGreen}{T_0}) } \sld^* \\
    \goal{ \mli{invoke}(\textcolor{red}{\mli{obj}(\mli{listfact}, \lpemptylist)}, \mli{buildlist}, [\mli{int}, \textcolor{blue}{A_0}], \textcolor{OliveGreen}{T_0}) } \sld^* \\
    \goal{ \mli{invoke}(\textcolor{red}{\mli{obj}(\mli{listfact}, \lpemptylist)}, \mli{buildlist}, [\mli{int}, \textcolor{blue}{A_1}], \textcolor{OliveGreen}{T_1}) } \sld^* \\
    \goal{ \mli{invoke}(\textcolor{red}{\mli{obj}(\mli{listfact}, \lpemptylist)}, \mli{buildlist}, [\mli{int}, \textcolor{blue}{A_i}], \textcolor{OliveGreen}{T_i}) } \sld \dotsb
\end{gather*}

However, as opposed to the previous example, in this case the derivation is \emph{not rational}. Indeed, at each step 
of the derivation a non-equivalent type is computed both for the accumulator and the returned value, since lists of different lengths have non-equivalent types. The following countably infinite set of equations defines the 
computed answer substitution associated with the whole derivation:
\begin{align*}
    A_0 &= \mli{obj}(\mli{elist}, \lpemptylist) & T_0 &= A_0 \lor T_1 \\
    A_1 &= \mli{obj}(\mli{nelist}, [\mli{head}\colon\mli{int}, \mli{tail}\colon A_0]) & T_1 &= A_1 \lor T_2 \\
    &\vdots & &\vdots \\
    A_i &= \mli{obj}(\mli{nelist}, [\mli{head}\colon\mli{int}, \mli{tail}\colon A_{i-1}]) & T_i &= A_i \lor T_{i+1} \\
    &\vdots & &\vdots
\end{align*}

After a closer look at the set above, we can deduce that the considered non rational derivation succeeds
because the set of equations above admits a solution, although such a solution involves non rational terms;
in particular, the type $T_0$ of the expression \mintinline{Java}{new List().buildList(42, new EList())} 
is non-rational; as a consequence, an inference engine based on Co-LP \cite{SimonMBG06} would fail
to compute a type, because no cycle can be detected in the derivation.

In order to solve this problem, we propose to use \emph{structural resolution} \cite{KomendantskayaJ15} as inference engine for abstract compilation. This new resolution method relies on a \emph{productivity} notion which is not limited to rational trees, thus it offers the possibility to exploit a more flexible inference engine to allow more expressive static type analysis through abstract compilation.

Starting from the goal above, after a finite number of derivation steps, structural resolution is able to compute the substitution\footnote{Depending on the implementation of structural resolution, the computed answer can be more or less precise, since type \(T_0\) could be ``unfolded'' more than once before the (first) answer is returned.} $T_0 = \mli{obj}(\mli{elist}, \lpemptylist) \lor T_1$, effectively solving the task of determining the type of the expression
\mintinline{Java}{new List().buildList(42, new EList())}. It works by noticing that the pattern $ \_ \lor \_$ is consumed by the terminating rewriting reductions, and is also infinitely produced in a chain of substitution reductions. Hence, $\lor$ serves as a constructor of the infinite data structure produced at infinity. 

The computed answer $T_0 = \mli{obj}(\mli{elist}, \lpemptylist) \lor T_1$ is only \emph{partial}, since variable $T_1$ is still ``unresolved''. However, structural resolution ensures that the non-rational term corresponding to the computed type $T_0$ can be incrementally unfolded for an \emph{arbitrary number} of steps: if needed, the substitution for \(T_1\) can be computed in a finite number of derivation steps, thus providing a better approximation of the type associated with \(T_0\). In this sense, the use of structural resolution in conjunction with abstract compilation gives rise to the implementation of a \emph{lazy} type inference procedure.

\section{Ensuring Universal Observability of Coinductive Logic Programming}\label{sec:prod}
\subsection{Universal Observability and Program Transformation}
In the previous section we discussed how structural resolution can be useful to handle (a class of) non-terminating derivations in finite time,  when derivations compute an infinite  irrational term. However, this new resolution method can be successfully employed only if logic programs are universally observable.

Logic programs resulting from abstract compilation are \emph{not} universally observable in the general case.

\begin{example}
\label{ex:subclass}
Consider for instance abstract compilation for an object-oriented language supporting nominal subtyping; the following three clauses should be generated for all source programs:
\begin{align*}
    \mli{subclass}(X, X) &\lpimpliedby \mli{class}(X) \\
    \mli{subclass}(X, \mli{object}) &\lpimpliedby \mli{class}(X) \\
    \mli{subclass}(X, Z) &\lpimpliedby \mli{extends}(X, Y) \lpand \mli{subclass}(Y, Z)
\end{align*}
The following infinite derivation shows that some logic programs obtained by abstract compilation are not universally observable:
\begin{dmath*}
\mli{subclass}(A, B) \rewriting
\mli{extends}(A, X) \lpand \mli{subclass}(X, B) \rewriting \mli{extends}(A, X) \lpand \mli{extends}(X, X') \lpand \mli{subclass}(X', B) \rewriting \dotsb
\end{dmath*}

Note that SLD-resolution would be able to find finite (i.e. inductive) proof for this query by unifying with the first or the second clause, whereas S-resolution's rewriting reductions get caught in an infinite loop by rewriting on the third clause. Thus for this example, S-resolution is also incomplete.
\end{example}

In order to fully exploit S-resolution, we present and formalise a \emph{transformation} of logic programs that ensures universal observability  of S-resolution reductions. Additionally, it ensures inductive completeness for inductive fragments of programs and observational productivity for coinductive fragments of programs. 
An extended version of this transformation has first been presented in \cite{FK16}.
Informally, given a clause $p(\overline{t}) \lpimpliedby q_1(\overline{t_1}) \lpand \dots \lpand q_n(\overline{t_n})$, the transformation adds extra argument
to all atoms, in such a way that the extra arguments reflect the clause structure:
$p(\overline{t},\kappa(\chi_1, \dots , \chi_n)) \lpimpliedby q_1(\overline{t_1},\chi_1) \lpand \dots \lpand q_n(\overline{t_n},\chi_n)$. Note that the extra term 
$\kappa(\chi_1, \ldots , \chi_n)$ can be intuitively read as: program clause $\kappa$ has $n$ atoms in its body.

The following definition formalises this translation for arbitrary logic programs. We use Greek letters for those parts of the logic program that are added in the transformation.

\begin{definition}[Productivity transformation]
\label{def:transf}
Given a logic program \(P\) we assume two sets \(\{\kappa_1, \kappa_2, \dotsc\}\) and \(\{\chi_1, \chi_2, \dotsc\}\) of distinct and fresh function symbols and variables, respectively. Then, \(\transf{P}\) is a new logic program defined by the following equations (for programs, clauses and atoms, respectively):
\begin{align*}
    \transf{P} &= \transf{\{C_1, \dotsc, C_n\}} = \{ \transf[\kappa_1]{C_1}, \dotsc, \transf[\kappa_n]{C_n} \} \\
    \transf[\kappa_i]{C_i} &= \transf[\kappa_i]{A \lpimpliedby B_1 \lpand \dots \lpand B_n} = \transf[\kappa_i(\chi_1, \dotsc, \chi_n)]{A} \lpimpliedby \transf[\chi_1]{B_1} \lpand \dots \lpand \transf[\chi_n]{B_n} \\
    \transf[\tau]{A} &= \transf[\tau]{p(t_1, \dotsc, t_n)} = p(t_1, \dotsc, t_n, \tau)
\end{align*}
Goal clauses have to be transformed as well in order to be resolved w.r.t.\ the transformed logic program:
\[ \transf{G} = \transf{A_1 \lpand \dots \lpand A_n} = \transf[\chi_1]{A_1} \lpand \dots \lpand \transf[\chi_n]{A_n} \]
\end{definition}

\begin{example} Consider the modified version of the program from example~\ref{ex:subclass}, where \(\kappa_i\) are distinct new function symbols and \(\chi\) is a new variable:
\begin{align*}
    \mli{subclass}(X, X, \kappa_1(\chi_1)) &\lpimpliedby \mli{class}(X, \chi_1) \\
    \mli{subclass}(X, \mli{object}, \kappa_2(\chi_1)) &\lpimpliedby \mli{class}(X, \chi_1) \\
    \mli{subclass}(X, Z, \kappa_3(\chi_1, \chi_2)) &\lpimpliedby \mli{extends}(X, Y, \chi_1) \lpand \mli{subclass}(Y, Z, \chi_2)
\end{align*}
It can be easily proved that this new version of the program \emph{is} universally observable, and the last argument provides the termination measure for rewriting reductions. In particular, $\mli{subclass}(A, B, \chi)  \not{\rewriting}$. Now we can have a one step substitution reduction to $\mli{subclass}(A, A, \kappa_1(\chi_1))$, and this subgoal will have one step (terminating) rewriting reduction to $\mli{class}(A, \chi)$, just as with SLD-resolution. 
\end{example}

As the above example shows, the program transformation guarantees that all  rewriting reductions terminate. It additionally has two different implications for inductive and coinductive programs. For inductive programs like the one from example~\ref{ex:subclass}, it allows completeness of derivations; and for coinductive programs  it allows lazy execution of S-resolution. 

\begin{example}
Consider this modified version of example \ref{ex:infinitelp}: 
\begin{align*}
    p(f(X), \kappa_1(\chi)) &\lpimpliedby p(X, \chi) \\
    q(X, \kappa_2(\chi)) &\lpimpliedby q(X, \chi) 
\end{align*}
The infinite S-resolution reductions for $q(X)$ now become both universally observable (all rewriting derivations terminate) and observationally productive:
\[
    q(X,\chi) \; \substitution^{\chi=\kappa_1(\chi')}  \; q(X,\kappa_1(\chi')) \; \rewriting \; q(X, \chi') \; \substitution^{\chi'=\kappa_1(\chi'')}\; q(X,\kappa_1(\chi'')) \; \rewriting \; q(X,\chi'') \;\dotsb 
\]
The S-resolution can now detect that the above reduction is productive in its second argument, i.e. an infinite term $\kappa_1(\kappa_1 \dots ))$ will be computed at infinity as a substitution to $\chi$ in the query $q(X,\chi)$.
\end{example}


The transformation above has a clear proof-relevant interpretation \cite{FK16}. Each rule gets a unique extra argument, and when a rule is applied, the instantiation of the last term of an atom can be understood as recording a proof evidence. In the end, the last term of each atom in the goal will be instantiated with a term encoding a \emph{proof} for the original atom \(A\) in the given program. For this reason, when we consider programs resulting from this transformation, we call terms in the last position inside atoms \emph{proof terms}.

We now present our original results showing that the productivity transformation of programs does not change their declarative -- inductive or coinductive -- semantics.

\begin{theorem}
\label{th:transfprod}
For any given logic program \(P\), \(\transf{P}\) is productive.
\begin{proof}
In order to show that \(\transf{P}\) is strongly normalizing, a decreasing measure on goal clauses needs to be established. Such a measure is the total number of function symbols in all the proof terms of the goal. It is easy to see that each clause \emph{reduces} this number by at least \(1\), since all the clauses in \(\transf{P}\) have the following shape:
\[ p(\dots, \kappa_i(\chi_1, \dots, \chi_m)) \lpimpliedby p_1(\dots, \chi_1) \lpand \dots \lpand p_m(\dots, \chi_m) \tag*{\qedhere} \]
\end{proof}
\end{theorem}

In light of its proof-relevant interpretation, the productivity transformation is expected not to change the intended meaning of the program since what it does is basically recording proof evidences. Indeed, we prove that \(\transf{}\) does not affect the declarative semantics of logic programs: the following result states that the transformation is sound and complete with respect to both the inductive and coinductive models.

\begin{theorem}[Soundness and completeness of the productivity transformation]
\label{th:soundcompl}
Given a logic program \(P\) and an atom \(A\), each of the following implications hold for some proof term \(\pi\):
\begin{align*}
    A \in M_P &\iff \transf[\pi]{A} \in M_{\transf{P}} && \text{(inductive soundness and completeness)} \\
    A \in M_P^{co} &\iff \transf[\pi]{A} \in M_{\transf{P}}^{co} && \text{(coinductive soundness and completeness)} 
\end{align*}
\begin{proof}
The full proof can be found in the Appendix \ref{sec:soundcompl}, and exploits the well-known identities 
$M_P = T_P \uparrow \omega$, and $M_P^{co} = T_P \downarrow \omega$ that hold for any logic program $P$ when the complete Herbrand base\footnote{For the first identity the standard inductive Herbrand base could be equivalently considered.} is considered \cite{Lloyd87}, and the two lemmas
stating the following claims:
$$
\begin{array}{l}
A \in T_P \uparrow n \iff \transf[\pi]{A} \in T_{\transf{P}} \uparrow n    \mbox{ for some proof term $\pi \in B_{\transf{P}}^{co}$}  \\
A \in T_P \downarrow n \iff \transf[\pi]{A} \in T_{\transf{P}} \downarrow n \mbox{ for some proof term $\pi \in B_{\transf{P}}^{co}$.} 
\end{array}
$$
Both lemmas are proved by induction over $n$; incidentally, the proofs show that $\pi$ actually coincides with a proof tree for $A\in M_P$ and $A\in M_P^{co}$, respectively.  
\end{proof}
\end{theorem}

In light of the results above, we can safely apply the transformation \(\transf{}\) to ensure productivity without changing the inductive and coinductive semantics of logic programs. This allows us to fruitfully exploit structural resolution as an inference engine together with abstract compilation.

\section{Conclusions}\label{sec:conclu}

Abstract compilation is a technique which uses logic programming for advanced static type analysis of object-oriented programs. 
To support analysis involving recursive types and programs, abstract compilation requires
to consider the coinductive interpretation of the generated logic programs. 
We have identified a class of recursive
methods for which the implementation of the inference engine based on Co-LP~\cite{SimonMBG06} does not work.
We  overcome this limitation by considering an inference engine based on structural resolution \cite{KPS12-2,KJS17}. 
S-resolution can only work if logic programs are universally observable; this property can be achieved by means of program transformation.
We have proposed a new translation scheme for abstract compilation  based on this transformation. 
We have
proved that such a transformation preserves the semantics of the programs generated by abstract compilation.

While these results show that there are cases where S-resolution works better than Co-LP
as inference engine for abstract compilation, we leave for further development the possibility
of extending such a claim to prove that structural resolution always leads to analysis results which, if not improved, are at least comparable to those obtained with Co-LP as inference engine of abstract compilation.
In particular, the recent results~\cite{Y17,KY17} show that it is possible to integrate Co-LP loop detection into S-resolution, and thus 
to identify regular patterns and infer regular terms like Co-LP does. For example, 
it is possible to infer the answer $X = cons(0,X)$ for a query of Example~\ref{ex:zeros} rather than giving a lazy answer  (\(X = cons(0,X'), X' = cons(0, X''), \dotsc\)).
A potential use of coinductive proofs has been investigated for type class inference in Haskell~\cite{FKSP16}, where irregular patterns arising in rewriting reductions (in the sense of Definition~\ref{def:Sres}) were considered. Future work will include investigation
on the existence of a unifying approach to different coinductive methods in logic programming, type inference and abstract compilation.

We are developing a prototype implementation~\cite{KSL17} based on a 
productivity checker for Logic Programming which provides an effective procedure for semi-deciding 
coinductive soundness of infinite S-resolution derivations, and, hence will result in a better assessment of the scalability of our proposed approach.

Another interesting direction for further development is the study of the ``completeness'' of our approach with respect to type inference problems.
This would allow us to better understand for which classes of programs the resolution (successfully) terminates in case of type inference queries; 
our prototype implementation could help us identify classes of real programs, by conducting experiments on, possibly simplified, code taken from
widely available open source projects.






\nocite{*} 
\bibliographystyle{eptcs}
\bibliography{bibliography}


\appendix
\section{Proofs of Soundness and Completeness of the Productivity Translation}
\label{sec:soundcompl}
\begin{definition}
Given a logic program \(P\), we use the following notations \cite{Lloyd87} for iterative application of the immediate consequence operator \(T_P\):
\begin{align*}
    T_P \uparrow 0 &= \emptyset & T_P \downarrow 0 &= B_P^{co} \\
    T_P \uparrow n+1 &= T_P(T_P \uparrow n) & T_P \downarrow n+1 &= T_P(T_P \downarrow n)
\end{align*}
\end{definition}

\begin{lemma}[Inductive soundness and completeness of \texttransf]
\label{lem:indsoundcompl}
Given a logic program \(P\), a (ground) atom \(A \in B_P\) and a number \(n \in \mathbb{N}\), the following implication holds for some proof term \(\pi \in B_{\transf{P}}\):
\[ A \in T_P \uparrow n \iff \transf[\pi]{A} \in T_{\transf{P}} \uparrow n \]
\begin{proof}
The proof goes by induction over \(n\). The base case is trivial since \(T_P \uparrow 0 = \emptyset\). When \(n > 0\) we prove the two implications as follows.
\begin{case}[\(\implies\)]
If \(n > 0\), since \(T_P \uparrow n = T_P(T_P \uparrow n-1)\), from the definition of \(T_P\) there must be a clause \(C = A' \lpimpliedby B_1 \lpand \dots \lpand B_m\) in \(P\) and a (grounding) substitution \(\sigma\) such that \(A = \sigma(A')\) and \(\{\sigma(B_1), \dots, \sigma(B_m)\} \subseteq T_P \uparrow n-1\). Since such a clause is in \(P\), its translation \(\transf[\kappa]{C}\) must be in \(\transf{P}\) for some constant \(\kappa\):
\[ \transf[\kappa]{C} = \transf[\kappa]{A' \lpimpliedby B_1 \lpand \dots \lpand B_m} = \transf[\kappa(\chi_1, \dotsc, \chi_m)]{A'} \lpimpliedby \transf[\chi_1]{B_1} \lpand \dots \lpand \transf[\chi_m]{B_m} \]
By inductive hypothesis, for some proof terms \(\pi_1, \dots, \pi_m\):
\[ \{\sigma(B_1), \dots, \sigma(B_m)\} \subseteq T_P \uparrow n-1 \implies \{\transf[\pi_1]{\sigma(B_1)}, \dots, \transf[\pi_m]{\sigma(B_m)}\} \subseteq T_{\transf{P}} \uparrow n-1 \]
Since \(\sigma\) is computed with respect to the original program \(P\) it does not change the proof terms \(\pi_i\), thus the substitution can be pushed out of the transformation:
\[ \{\sigma(\transf[\pi_1]{B_1}), \dots, \sigma(\transf[\pi_m]{B_m})\} \subseteq T_{\transf{P}} \uparrow n-1 \]
Through one application of \(T_{\transf{P}}\) we can apply the clause \(\transf[\kappa]{C}\) to the atoms above with the substitution mapping every variable \(\chi_i\) to the proof term \(\pi_i\):
\[ \sigma(\transf[\kappa(\pi_1, \dots, \pi_m)]{A'}) \in T_{\transf{P}} \uparrow n \]
Finally, we conclude by moving the substitution \(\sigma\) again, recalling that \(\sigma(A') = A\) and \(\kappa(\pi_1, \dots, \pi_m)\) is ground:
\[ \sigma(\transf[\kappa(\pi_1, \dots, \pi_m)]{A'}) = \transf[\sigma(\kappa(\pi_1, \dots, \pi_m))]{\sigma(A')} = \transf[\kappa(\pi_1, \dots, \pi_m)]{A} \in T_{\transf{P}} \uparrow n \]
\end{case}
\begin{case}[\(\impliedby\)]
The opposite implication can be proved in quite a similar way. From the definitions of \(T_P\) and \(T_P \uparrow n\) we know the two following clauses are in \(P\) and \(\transf{P}\), respectively:
\begin{gather*}
    C_i = A' \lpimpliedby B_1 \lpand \dots \lpand B_m \in P \\
    \transf[\kappa_i]{C_i} = \transf[\kappa_i(\chi_1, \dots, \chi_m)]{A'} \lpimpliedby \transf[\chi_1]{B_1} \lpand \dots \lpand \transf[\chi_m]{B_m} \in \transf{P}
\end{gather*}
and, also, there is a substitution \(\sigma\) such that \(A = \sigma(A')\) and, for \(i=1,\dots,n\):
\[ \sigma(\transf[\chi_i]{B_i}) = \transf[\sigma(\chi_i)]{\sigma(B_i)} \in T_{\transf{P}} \uparrow n-1 \]
Then, by inductive hypothesis:
\[ \sigma(B_i) \in T_P \uparrow n-1 \]
Finally, applying \(T_P\) to \(\{B_1,\dots,B_m\}\) with clause \(C_i\) and substitution \(\sigma\), we can conclude that \(\sigma(A') = A \in T_P \uparrow n\).
\end{case}
\end{proof}
\end{lemma}

\begin{lemma}[Coinductive soundness and completeness of \texttransf]
\label{lem:cosoundcompl}
Given a logic program \(P\), a (ground) atom \(A \in B_P^{co}\) and a number \(n \in \mathbb{N}\), the following implication holds for some proof term \(\pi \in B_{\transf{P}}^{co}\):
\[ A \in T_P \downarrow n \iff \transf[\pi]{A} \in T_{\transf{P}} \downarrow n \]
(note that both \(A\) and \(\pi\) could be infinite)
\begin{proof}
The proof goes by induction over \(n\). The base case is trivial since \(T_P \downarrow 0 = B_P^{co} \) by definition. The inductive case is entirely similar to the proof of lemma \ref{lem:indsoundcompl}, the only difference being the use of \(T_P \downarrow n\) rather than \(T_P \uparrow n\).
\end{proof}
\end{lemma}

\paragraph{Proof of Theorem \ref{th:soundcompl}} (\emph{Soundness and completeness of the productivity transformation}).
The following well-known results \cite{Lloyd87} about the least and greatest fixed points hold for any logic program \(P\) when the complete Herbrand base is considered:
\begin{align*}
    M_P &= T_P \uparrow \omega \\
    M_P^{co} &= T_P \downarrow \omega
\end{align*}
The theorem therefore directly follows from lemmas \ref{lem:indsoundcompl} and \ref{lem:cosoundcompl}, that proved the strict relation between \(T_P\) and \(T_{\transf{P}}\). \null\hfill\qedsymbol

\end{document}